\documentclass[letter,twocolumn]{jpsj2} %% two-column layout
\title{Electronic Structure and Electron Correlation in LaFeAsO$_{1-x}$F$_x$ and LaFePO$_{1-x}$F$_x$}

\author{
Walid \textsc{MALAEB}$^{1}$\thanks{malaeb@wyvern.phys.s.u-tokyo.ac.jp},
Teppei \textsc{YOSHIDA}$^{2}$,
Takashi \textsc{KATAOKA}$^{1}$,
Atsushi \textsc{FUJIMORI}$^{1}$,
Masato \textsc{KUBOTA}$^{3}$,
Kanta \textsc{ONO}$^{3}$,
Hidetomo \textsc{USUI}$^{4}$,
Kazuhiko \textsc{KUROKI}$^{4}$,
Ryotaro \textsc{ARITA}$^{5}$,
Hideo \textsc{AOKI}$^{2}$,\\
Yoichi \textsc{KAMIHARA}$^{6}$,
Masahiro \textsc{HIRANO}$^{6,7}$,
and Hideo \textsc{HOSONO}$^{6,7}$}

\inst{
$^{1}$Department of Complexity Science and Engineering and Department of Physics, University of Tokyo, Bunkyo-ku, Tokyo 113-0033\\
$^{2}$Department of Physics, University of Tokyo, Bunkyo-ku, Tokyo 113-0033 \\
$^{3}$Institute for Material Structure Science, High Energy Accelerator Research Organization, Tsukuba, Ibaraki 305-0801\\
$^{4}$Department of Applied Physics and Chemistry, The University of Electro-Communications, Chofu, Tokyo 182-8585\\
$^{5}$Department of Applied Physics, University of Tokyo, Bunkyo-ku, Tokyo 113-8561\\
$^{6}$ERATO-SORST, JST, in Frontier Research Center, Tokyo Institute of Technology, Mail Box S2-13,259 Nagatsuta, Midori-ku, Yokohama 226-8503\\
$^{7}$Frontier Research Center, Tokyo Institute of Technology,
Mail Box S2-13, 4259 Nagatsuta, Midori-ku, Yokohama 226-8503}

\abst{Photoemission spectroscopy is used to investigate the
electronic structure of the newly discovered iron-based
superconductors LaFeAsO$_{1-x}$F$_x$ and LaFePO$_{1-x}$F$_x$. Line
shapes of the Fe $2p$ core-level spectra suggest an itinerant
character of Fe $3d$ electrons. The valence-band spectra are
generally consistent with band-structure calculations except for
the shifts of Fe $3d$-derived peaks toward the Fermi level. From
spectra taken in the Fe $3p \to 3d$ core-absorption region, we
have obtained the experimental Fe $3d$ partial density of states,
and explained it in terms of a band-structure  calculation  with a
phenomenological self-energy correction, yielding a mass
renormalization factor of $\sim <$ 2.}

\kword{ LaFeAsO, electronic structure, photoemission spectroscopy }

\begin{document}
\maketitle
There is surging interest toward the high-$T_c$
superconductivity recently reported in the iron-based compound
LaFeAsO$_{1-x}$F$_x$ \cite{ref1}, which has been followed by
reports on other compounds belonging to the same family,
$Ln$FeAsO$_{1-x}$F$_x$ ($Ln$ = La, Ce, Pr, Nd, Sm), with $T_c$ up
to $\sim$55 K in SmFeAsO$_{1-x}$F$_x$ \cite{ref2}. The latter
$T_c$ is the highest to date apart from the high-$T_c$ cuprates.
LaFeAsO$_{1-x}$F$_x$ has LaO and FeAs layers alternatingly stacked
along the $c$-axis, which renders the compound highly
two-dimensional physical properties similar to the cuprates. The
parent material LaFeAsO is a semiconductor or a bad metal while
the system shows superconductivity with fluorine doping, which is
believed to induce electrons into the conducting FeAs layer
\cite{ref1}.

\begin{figure}[tb]
\begin{center}
\includegraphics[width=90mm]{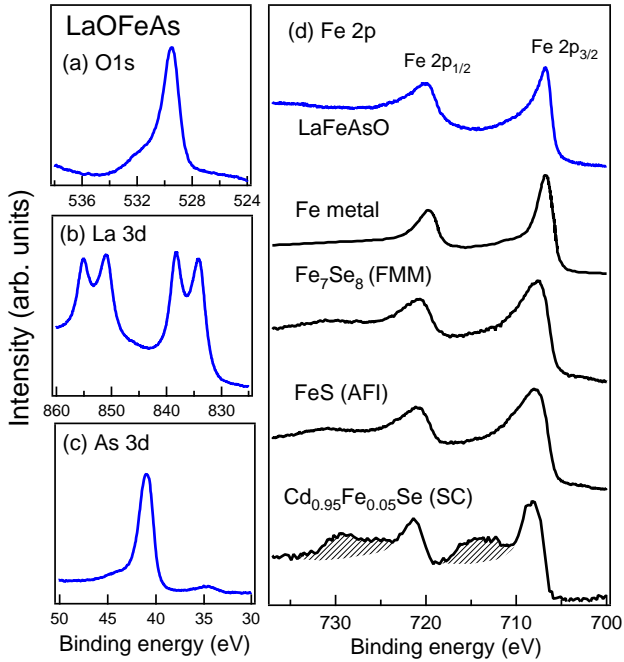}
\end{center}
\caption{Core-level XPS spectra of LaFeAsO: (a) O $1s$, (b) La
$3d$, (c) As $3d$, (d) Fe $2p$. The Fe $2p$ spectrum is compared
with those of Cd$_{0.95}$Fe$_{0.05}$Se (semiconductor), FeS
(antiferromagnetic insulator), Fe$_7$Se$_8$ (ferromagnetic metal)
\cite{ref9} and Fe metal \cite{ref9'}.} \label{f1}
\end{figure}

Although the superconducting gap and pseudogap have been studied
by ultra-high resolution photoemission spectroscopy
\cite{ref3,ref4,ref3'} along with a recent report of
angle-resolved photoemission spectroscopy (ARPES) on single
crystals \cite{ref4''}, basic knowledge of the electronic
structure of the iron-based superconductors that can distinguish
themselves from those of other superconductors has yet to come.
For example, it is well known that strong electron correlation
plays a major role in the high-$T_c$ cuprates, while it is not
clear to what extent this applies to LaFeAsO$_{1-x}$F$_x$. Also,
strong $p$-$d$ hybridization exists in the cuprates, whereas no
clear idea on this is known for LaFeAsO$_{1-x}$F$_x$. Some
theoretical works imply the importance of electron correlations
\cite{ref5}, and their effects on the unconventional
superconductivity \cite{ref6,ref6',ref7}. Photoemission
spectroscopy (PES) is one of the most powerful tools to study the
electronic structure of solids and electron correlation effects.
In the present work, we have used PES to investigate the
core-level and valence-band spectra of LaFeAsO and LaFePOF.

Polycrystals of LaFeAsO$_{1-x}$F$_x$ ($x=$ 0, 0.06) and
LaFePO$_{1-x}$F$_x$ ($x=$ 0.06) were synthesized as described
elsewhere \cite{ref1,ref8}. The x-ray photoemission spectroscopy
(XPS) measurements were performed using an Mg $K\alpha$  source
($h\nu=$ 1253.6 eV) at 15 K.  The samples were repeatedly scraped
with a diamond file to obtain clean surfaces. High-resolution
ultraviolet photoemission spectroscopy measurements were performed
at beamline 28A of Photon Factory, KEK, with the energy resolution
of $\sim$20 meV at 15 K. The samples were fractured \textit{in
situ} in an ultra-high vacuum below 1$\times$10$^{-10}$ Torr.
Theoretical partial density of states has been calculated as
follows. We first obtained the band structure in the same way as
Kuroki {\it et al.} \cite{ref6}. Then, we constructed the
maximally localized Wannier functions (MLWFs) for the Fe $3d$, As
$4p$/P $3p$, and O $2p$ orbitals using a code developed by Mostofi
{\it et al.} \cite{ref8''}. and calculated Green's function to
obtain the spectral function for each MLW.

Figure \ref{f1} shows the O $1s$, La $3d$, As $3d$ and Fe $2p$
core-level spectra of LaFeAsO. The single O $1s$ peak with a
largely diminished high-binding energy shoulder (Fig.~\ref{f1}(a))
reflects the cleanliness of the sample surface. Nearly identical
spectra were obtained for F-doped samples
LaFeAsO$_{0.94}$F$_{0.06}$ and LaFePO$_{0.94}$F$_{0.06}$ (not
shown). In Fig.~\ref{f1}(d), the Fe $2p$ spectrum of LaFeAsO is
compared with those of other iron compounds in the literature
\cite{ref9,ref9'}. The satellites observed in the Fe $2p$ spectrum
of the diluted magnetic semiconductor Cd$_{0.95}$Fe$_{0.05}$Se
(shaded area) reflects the localized character of the Fe $3d$
electrons in this compound \cite{ref9}. The spectrum of the
antiferromagnetic insulator FeS \cite{ref9} is broad and has a
high ``background" intensity on the higher binding energy side of
the Fe $2p_{3/2}$ peak. On the other hand, the Fe $2p$ core-level
spectrum of LaFeAsO shows no such satellites nor high
binding-energy background intensity, while the Fe $2p_{3/2}$ peak
itself is as sharp as that of elemental Fe \cite{ref9'}. These
observations indicate an itinerant nature of the Fe $3d$ electrons
in LaFeAsO. This is consistent with an NMR result on LaFeAsO,
according to which the system shows itinerant antiferromagnetism
with $T_{\rm N}\sim$142 K and antiferromagnetic fluctuations in
F-doped compounds \cite{ref10}.

\begin{figure}[tb]
\begin{center}
\includegraphics[width=95mm]{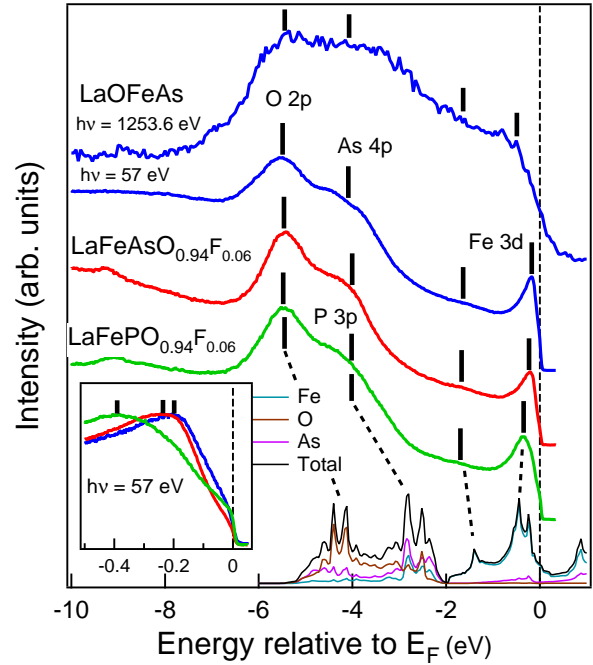}
\end{center}
\caption{Valence-band photoemission spectra of
LaFeAsO$_{1-x}$F$_x$ ($x=$ 0, 0.06) and LaFePO$_{0.94}$F$_{0.06}$
and their comparison with band-structure calculation. Vertical
bars mark main features observed in the spectra. The inset
presents the near-$E_{\rm F}$ spectra} \label{f2}
\end{figure}

Figure \ref{f2} presents the valence-band spectra of
LaFeAsO$_{1-x}$F$_x$ ($x=$ 0, 0.06) and LaFePO$_{0.94}$F$_{0.06}$.
Main features in the valence band common to the three compounds
are a sharp peak near the Fermi level ($E_{\rm F}$), a weak
structure at $\sim$-1.5 eV, a shoulder at $\sim$-4 eV and a broad
peak at $\sim$-5.5 eV, consistent with the previous report on
LaFeAsO$_{1-x}$F$_x$ \cite{ref3}. Similar features are observed in
the valence band measured by XPS displayed also in the same
figure. For comparison, the band-structure calculation result is
displayed at the bottom of Fig.~\ref{f2}. As in the previous
calculation \cite{ref10'}, the present result predicts a peak in
the density of states (DOS) near the Femi level, where the main
contribution comes from the Fe $3d$ states. Contributions between
-2 and -5 eV are mainly from As $4p$/P $3p$ (at $\sim$-3 eV) and O
$2p$ (at $\sim$-4.5 eV). We then attribute the peaks near $E_{\rm
F}$ in the photoemission spectra to Fe $3d$ states, while the
shoulder around -4 eV and the peak around -5.5 eV to As $4p$/P
$3p$ and O $2p$ states, respectively. Although the experimental
data agree qualitatively well with the calculation, some
differences are observed where the peak positions near $E_{\rm F}$
and the other peaks observed in experiment occur at somewhat lower
and higher binding energies, respectively, than predicted by the
calculation.

In a blowup near $E_{\rm F}$ (inset of Fig.~\ref{f2}), one notices
that the peaks are shifted towards higher binding energies with F
doping in LaFeAsO$_{1-x}$F$_x$. The shift can be explained as a
chemical potential shift due to the electron doping. As a result,
the intensity of the spectra at $E_{\rm F}$ decreases with F
doping. Also, the Fe $3d$ peak for LaFePO$_{1-x}$F$_x$ is located
at higher binding energies and is broad as compared to that of
LaFeAsO$_{0.94}$F$_{0.06}$. The broadness of the Fe $3d$ band of
LaFePO$_{0.94}$F$_{0.06}$ as compared to
LaFeAsO$_{0.94}$F$_{0.06}$ has been predicted by band-structure
calculations, and can be attributed to the difference in the ionic
radii of As and P, which makes the Fe-P distance (0.229 nm)
shorter than Fe-As distance (0.241 nm) \cite{ref1,ref8}, resulting
in a larger value of the hopping parameter ($t$) for LaFePO than
that of LaFeAsO. Also, it is noted that LaFePO$_{0.94}$F$_{0.06}$
has a higher intensity at $E_{\rm F}$ than
LaFeAsO$_{0.94}$F$_{0.06}$. This means that, although
LaFeAsO$_{0.94}$F$_{0.06}$ shows superconductivity at $T_c \simeq$
26 K, its DOS at $E_{\rm F}$ are smaller than those observed for
non-superconducting LaFeAsO and LaFePO$_{0.94}$F$_{0.06}$ with
$T_c =$ 3-5 K. This suggests that effects other than the DOS at
$E_{\rm F}$, such as Fermi surface shapes and coupling to boson
excitations, may be at work for the superconductivity.

\begin{figure}[tb]
\begin{center}
\includegraphics[width=90mm]{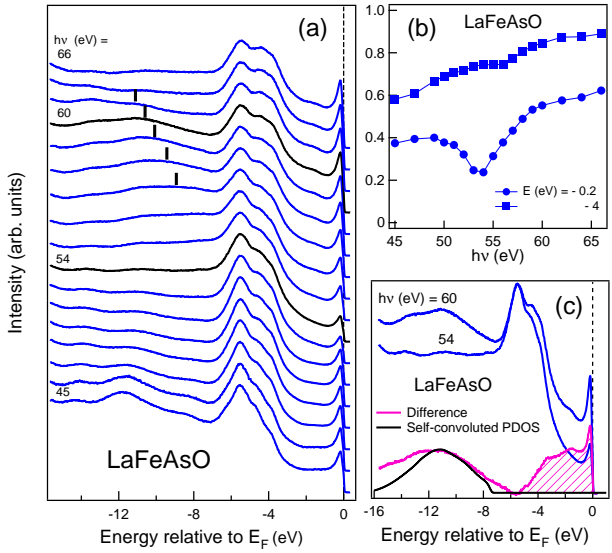}
\end{center}
\caption{Valence-band photoemission spectra of LaFeAsO in Fe $3p
\to 3d$ core absorption region. (a) A series of photoemission
spectra for various photon energies. Vertical bars indicate the
$M_{2,3}M_{4,5}M_{4,5}$ Auger peak. (b) Plot of the photoemission
intensities at $E =$ -0.2 eV and -4.0 eV as functions of photon
energy.  (c) Comparison of the self-convolution of the Fe $3d$
partial density of states (PDOS) (shaded part of the difference
spectrum) with the Auger-electron spectrum.} \label{f3}
\end{figure}

The valence-band spectra of LaFeAsO taken at various photon
energies in the Fe $3p\to 3d$ core excitation region are shown in
Fig.~\ref{f3}(a). Here the spectra have been normalized to the O
$2p$ peak intensity at -5.5 eV. One can see that the intensity of
the near-$E_{\rm F}$ peaks and the -4 eV shoulder show dramatic
photon energy dependence: They exhibit an increase from $h\nu
\sim$ 54 eV to $h\nu\sim$ 60 eV. The $h\nu$-dependence plotted in
Fig.~\ref{f3}(b) is indicative of the Fe $3p \to 3d$ resonance,
and reconfirms that the near-$E_{\rm F}$ states are mainly Fe $3d$
states and that the -4 eV shoulder representing the As $4p$ band
is significantly hybridized with Fe $3d$.

The high binding energy part ($\sim <$ -6 eV) of the spectra in
Fig.~\ref{f3}(a) also shows a characteristic dependence on photon
energy. The broad feature which is shifted to higher binding
energies and grows with increasing photon energy, marked by
vertical bars in Fig.~\ref{f3}(a), is due to Fe
$M_{2,3}M_{4,5}M_{4,5}$ (Fe $3p$-$3d$-$3d$) Auger-electron
emission. In the final state of the Auger-electron emission, two
holes are left in the Fe $3d$ band, making Auger electron
spectroscopy a good probe to investigate the on-site Coulomb
energy \cite{ref11}. The good agreement between the Auger spectrum
and the self-convolution of the Fe $3d$ PDOS indicates that $U \ll
2W$, where $U$ is Fe $3d$ on-site Coulomb energy and $W$ is the Fe
$3d$ band width \cite{ref11}, and therefore most likely $U < W$,
confirming that the band description of the Fe $3d$ band is a good
starting point to understand the electronic properties of
LaFeAsO$_{1-x}$F$_x$.

\begin{figure}[tb]
\begin{center}
\includegraphics[width=90mm]{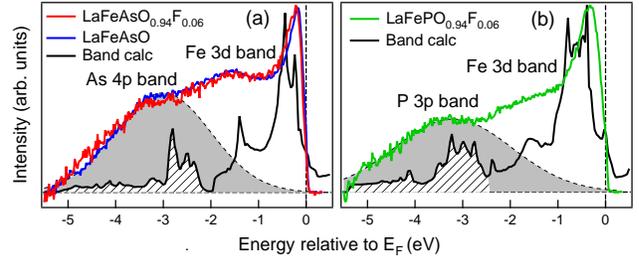}
\end{center}
\caption{Fe 3d PDOS of LaFeAsO$_{1-x}$F$_x$ (a) and
LaFePO$_{0.94}$F$_{0.06}$ (b) determined by subtracting the
off-resonance spectra (taken at $h\nu =$ 54 eV) from the
on-resonance ones ($h\nu =$ 60 eV) (see Fig.~\ref{f3}(c)). In
order to isolate the Fe $3d$ band, the As $4p$/P $3p$ band assumed
to be a Gaussian has been subtracted.} \label{f4}
\end{figure}

\begin{figure}[tb]
\begin{center}
\includegraphics[width=90mm]{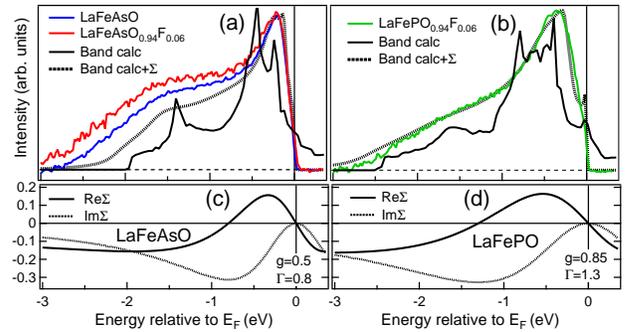}
\end{center}
\caption{Results of self-energy corrections to the Fe $3d$ band
for LaFeAsO$_{1-x}$F$_x$ (a) and LaFePO$_{0.94}$F$_{0.06}$ (b).
(c), (d): Energy dependence of the real and imaginary parts of the
empirically determined self-energy $\Sigma(\omega)$.} \label{f5}
\end{figure}

In order to deduce the experimental Fe $3d$ PDOS, we have
subtracted the off-resonance spectra (taken at $h\nu=$54 eV) from
the on-resonance spectra ($h\nu=$60 eV) as shown in
Fig.~\ref{f3}(c). The Fe $3d$ PDOS thus obtained shown in
Fig.~\ref{f4} indicates that the near-$E_{\rm F}$ peak and the
weak peak at $\sim$-1.5 eV corresponding to Fe $3d$ bands survive
and that a broad peak corresponding to the As $4p$/P $3p$ band
appears in the range -(3-4) eV for every compound. To extract the
Fe $3d$ band, the As $4p$/P $3p$ band is approximated by a
Gaussian and has been subtracted from the Fe $3d$ PDOS, leaving
the Fe $3d$-band part of the Fe $3d$ PDOS. In order to account for
the deviations of the experimental Fe $3d$ PDOS from the
calculated one, we have applied a self-energy correction to the
band calculation result. We take an empirical approach, where we
assume that electron correlation gives rise to a self energy
$\Sigma(\omega)$, where $\Sigma$ is assumed to be
$\omega$-dependent but momentum independent, to retain the
Fermi-liquid properties ($\Sigma(\omega)\sim -a \omega- {\rm
i}b\omega^2$ in the vicinity of $E_{\rm F}$) and to satisfy the
Kramers-Kronig relation.  Here, we take a simple analytical form
$\Sigma(\omega) = -g\omega/(\omega+{\rm i}\Gamma)^2$, for which
Im$\Sigma$ and Re$\Sigma$ are shown in Fig.~\ref{f4}(c) and (d),
and the parameters are fitted to reproduce the experimental
spectra, as has previously done for Fe chalcogenides \cite{ref9}.
While the iron pnicties are theoretically conceived as multiband
systems \cite{ref6}, this treatment amount to assuming that
orbital dependence in the self-energy can be ignored. The single
particle spectral DOS, $\rho(\omega)$, is thus given by
\begin{eqnarray}
\rho(\omega)=-\frac{1}{\pi}\int d\epsilon N_b(\epsilon){\rm Im}\frac{1}{\omega-\epsilon-\Sigma(\omega)},
\end{eqnarray}
where $N_b(\epsilon)$ is the Fe $3d$ PDOS given by the
band-structure calculation. The $\rho(\omega)$ thus obtained
(Fig.~\ref{f4}(a) and (b)) is seen to exhibit better agreement
with experiment.  The self-energy gives rise to a mass
enhancement, $m^*/m_b = 1- \partial{\rm
Re}\Sigma(\omega)/\partial\omega = 1+g/\Gamma^2$, where $m_b$ is
the bare band mass and $m^*$ the enhanced mass at $E_{\rm F}$. We
obtain $m^*/m_b \simeq 1.8$ for LaFeAsO$_{1-x}$F$_x$, and $m^*/m_b
\simeq 1.5$ for LaFePO$_{0.94}$F$_{0.06}$. These values are in the
same range as the experimental values deduced from the Seebeck
coefficient and thermal conductivity \cite{ref12}. For more
precise discussions, more elaborate analyses of ARPES data with a
band-dependent self-energy will be necessary in future studies.

In conclusion, we have investigated the electronic structure of
LaFeAsO$_{1-x}$F$_x$ and LaFePO$_{1-x}$F$_x$ by photoemission
spectroscopy. The Fe $2p$ core-level spectra indicate an itinerant
behavior rather than strongly correlated one. The valence-band
spectra are consistent with the band-structure calculations, and
show that Fe $3d$ states are dominant near the Fermi level.
Existence of a moderate electron correlation and $p$-$d$
hybridization have been demonstrated through the renormalization
of the Fe $3d$ band.

The authors wish to acknowledge Y. Ishida for useful information
on experimental details, and M. Kobayashi for informative
discussions. This work was supported by a Grant-in-Aid for
Scientific Research in Priority Area ``Invention of Anomalous
Quantum Materials" from MEXT and by CREST, Japan Science and
Technology Agency. W.M. is thankful to MEXT for a financial
support. Experiment at Photon Factory was approved by the Photon
Factory Program Advisory Committee (Proposal No. 2006S2-001).

\end{document}